\documentclass[letterpaper, 10 pt, conference]{ieeeconf}  

\IEEEoverridecommandlockouts                              
\usepackage{cite}
\usepackage{graphicx} 
\usepackage{epsfig} 
\usepackage{amsmath} 
\usepackage{amssymb}  
\usepackage[hyphens]{url}
\usepackage{color}
\usepackage{algorithm,algorithmic}
\title{
\LARGE \bf Transfer Operator Theoretic Framework for Monitoring Building Indoor Environment in Uncertain Operating Conditions}

\author{Himanshu Sharma$^{1}$, Anthony D. Fontanini$^{2}$, Umesh Vaidya$^{3}$, Baskar Ganapathysubramanian$^{4}$
\thanks{$^{1}$Graduate student, Dept. of Mechanical Engineering, Iowa State University, IA, Ames,50010 
        {\tt\small hsharma@iastate.edu}}%
\thanks{$^{2}$Fraunhofer CSE, Boston, MA 02210, USA
        {\tt\small anthony.fontanini@gmail.com }}%
\thanks{$^{3}$Faculty, Dept. of Electrical and Computer Engineering, Iowa State University, IA, Ames, 50010
        {\tt\small ugvaidya@gmail.com}}%
\thanks{$^{4}$Faculty, Dept. of Mechanical Engineering, Iowa State University, IA, Ames, 50010
        {\tt\small baskarg@iastate.edu}}%
}

\begin{document}

\maketitle
\thispagestyle{empty}
\pagestyle{empty}

\begin{abstract}
Dynamical system-based linear transfer Perron-Frobenius (P-F) operator framework is developed to address analysis and design problems in the building system. In particular, the problems of fast contaminant propagation and optimal placement of sensors in uncertain operating conditions of indoor building environment are addressed. The linear nature of transfer P-F operator is exploited to develop a computationally efficient numerical scheme based on the finite dimensional approximation of P-F operator for fast propagation of contaminants. The proposed scheme is an order of magnitude faster than existing methods that rely on simulation of an advection-diffusion partial differential equation for contaminant transport. Furthermore, the system-theoretic notion of observability gramian is generalized to nonlinear flow fields using the transfer P-F operator. This developed notion of observability gramian for nonlinear flow field combined with the finite dimensional approximation of P-F operator is used to provide a systematic procedure for optimal placement of sensors under uncertain operating conditions. Simulation results are presented to demonstrate the applicability of the developed framework on the IEA-annex 2D benchmark problem. 

\end{abstract}

\section{INTRODUCTION}
Indoor air quality sensing is important in maintaining healthy environments in buildings. A recent survey finding from the American Lung Association's (ALA) "state of the air 2017" reports that nearly 40\% of the American population live in counties which are under unhealthy ozone or particulate pollutant levels \cite{ALA2017}. A study from WHO \cite{WHO2014} has also investigated the significance of indoor air pollution on health. The result from WHO states that indoor air pollution was the primary cause of nearly 4.3 million deaths globally in 2012. An individual spends most of their time in the indoor environment, and therefore an unhealthy indoor environment can be more dangerous than the outdoor pollutants. Infants and older adults are subjected to greater risk of the indoor pollutants such as harsh cleaning chemicals(VOCs), pollen, smoke, and dust. The removal of these contaminants is essential to ensure clean breathing environment for the occupants.

Additionally, it is also important to maintain air quality in public spaces where the risk of transmission of infectious diseases (TID) like tuberculosis, influenza, SARS is higher. This is possible by careful design of a sensor network which can identify and localize contaminants. Such networks can also be critical for indoor security in case of extreme events such as chemical and/or biological warfare (CBW). 

\indent Overall, the strategy for ensuring a healthy indoor environment can be divided into three stages (1) \textit{ risk assessment and prevention}, (2) \textit{ source identification}, and (3) \textit{ post-identification response strategies}. The optimal placement of sensor placement and pollutant estimator design come under the category of risk assessment. For optimal sensor placement the methods available in the literature can be broadly classified into three categories (1) \textit{Engineering \& Heuristics} (2) \textit{Optimization \& Inverse modelling} (3) \textit{Dynamics Systems Approach}. 
The engineering/heuristic-based approach is fast, but the challenges of this approach includes; lack of formal procedures to design a response time, may result in full or partial coverage for monitoring the indoor environment, does not account for uncertainty
and faces challenges to address sensor placement in multiple connected rooms or zones \cite{Liu2009}. To overcome the problem of this approach, optimization and inverse methods have recently gained attention \cite{Liu2009a,Mazumdar2008,Zhang2007,Chen2008,zhou2009optimization}. However, these methods are not as widely used as expected due to the challenges of computational cost and the complexity of solving partial differential equations for multiple release scenarios.\\
\indent In this paper, we propose a framework based on linear transfer Perron-Frobenius (P-F) operator to address the problem of contaminant propagation and optimal placement in the building system. The basic idea behind the proposed framework is to replace the nonlinear evolution of contaminant under a fluid flow field with a linear albeit infinite dimensional evolution of densities or measures. The finite dimensional approximation of P-F operator arises as a Markov matrix. Contaminant propagation in finite dimensions then corresponds to simple matrix-vector products and hence can be performed relatively quickly \cite{Chen, Chen2015, Fontanini2015, Fontanini2017}. Most of the computational burden is involved in the construction of the finite dimensional approximation of P-F operator, but this can be performed off-line beforehand. The linear nature of the P-F operator is further exploited to extend the notion of observability gramian for a nonlinear fluid flow field for accomplishing optimal sensor placement. A systematic procedure based on the maximization of observability to ensure observability in the entire indoor environment is proposed for optimal sensor placement. Furthermore, the incorporation of various sources of uncertainty such as weather conditions, occupancy, interior design and construction materials is conceptually (and implementation wise) easy to incorporate in the framework. This allows us to extend the results on optimal sensor placement and fast contaminant transport to building system dynamics with uncertainty. 
The outline for the paper is as follows. We discuss the method for constructing the transfer P-F operator for the contaminant transport and the sensor placement in the deterministic setting in section-\ref{sec:TransOper}, The extension of the P-F approach for the uncertain operating conditions is discussed in section-\ref{sec:UncertainOper}. The results of the developed P-F operator-based method on IEA-annex 2D benchmark problem are presented in section-\ref{sec:Results}.
\section{Transfer P-F operator for contaminant transport and optimal placement}\label{sec:TransOper}
\begin{figure}[h]
\centering
\includegraphics[scale=0.45]{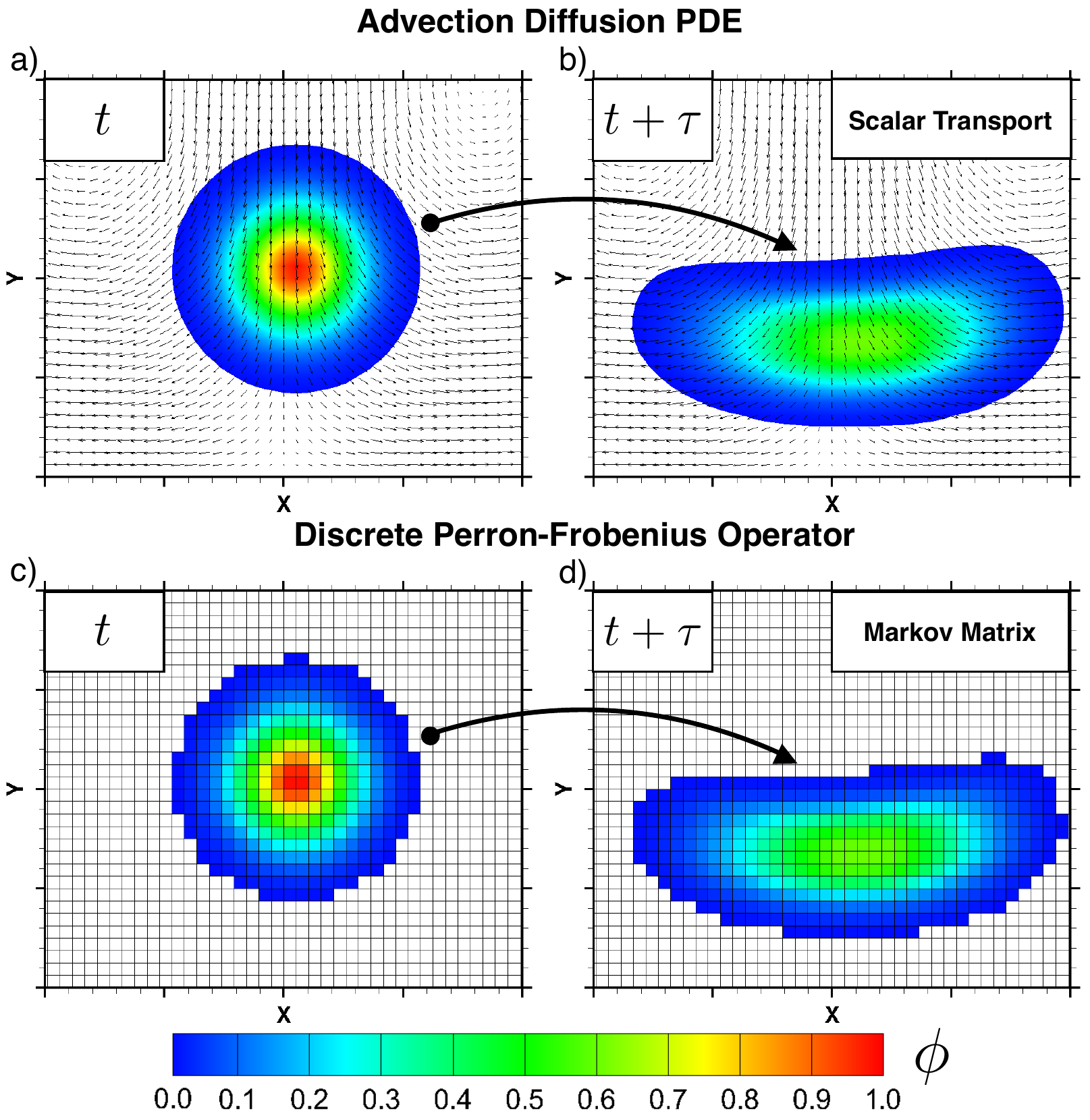}
\caption{For the given velocity field (a)-(b)Shows the contaminant transport using the scalar transport equation-\ref{Eq:1}(c)-(d)Shows the discrete PF-operator based scalar transport \cite{Fontanini2016}}.
\label{Fig:Euler_transp}
\vspace*{-1em}
\end{figure}

Consider the following advection-diffusion partial differential equation (PDE) with source term and output measurement.
\begin{eqnarray}
&&\frac{\partial \Phi}{\partial t} + \nabla(U\Phi) +\nabla^2(D\Phi) = S_{\Phi} \nonumber\\
&&y=\chi_{A_k}(x)\Phi,\;\;k=1,\ldots,p
\label{Eq:1}
\end{eqnarray} 
where $\Phi(x,t)$ is the scalar contaminant density at time $t$, $x\in X\subset{R}^n$ is the physical space,  $D$ is the diffusion constant and $S_{\Phi}$ is the source term in Eq. (\ref{Eq:1}) 
The velocity flow field $U$ can be steady or can be a function of time. The scalar contaminant is propagated by the given airflow field, $U$. The flow field can be generated experimentally, or computationally using computational fluid dynamics (CFD). $\chi_{A_k}(x)$ for $k=1,\ldots,p$ denotes the indicator function for set $A_k\subset X$ and corresponds to the location of $p$ sensors.  Note that the advection-diffusion PDE (\ref{Eq:1}) describe linear evolution for the propagation of contaminant $\Phi$. This continuous time linear evolution can be replaced in discrete-time using linear transfer P-F operator. In the absence of source term in Eq. (\ref{Eq:1}), the discrete-time evolution of advection-diffusion PDE is described by following P-F operator
\begin{eqnarray}
[{\mathbb{P}\mu}](A)=\int_X p(x,A)d\mu(x).
\end{eqnarray}
where $\mu\in {\cal M}(X)$ the space of real-valued measure and $p(x,A)$ is stochastic transition function and describes the transition probability from point $x$ to set $A\subset X$. To compute the finite dimensional approximation of the P-F operator we only need the information about the nonlinear flow field $U$ and the diffusion coefficient $D$ as is described in the following section. The example of this is shown in Fig.\ref{Fig:Euler_transp}, where the velocity field is shown by vectors and the color contours as the initial concentration.
\subsection{Finite dimensional approximation of P-F operator and contaminant propagation}\label{sec:ConstructPF}
\begin{figure}[h]
\centering
\includegraphics[scale=0.45,width=0.78\linewidth]{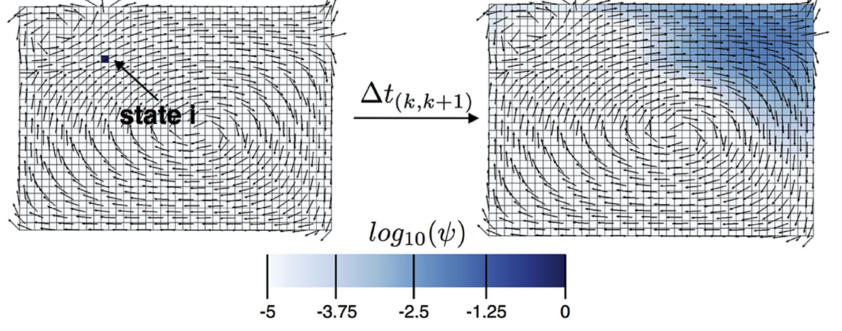}
\caption{The Eulerian approach of constructing Markov matrix for calculating single row $i$ of P. The $D_{k}$ state is initialized with 1.0 and then the advection-diffusion equation are solved to compute the evolution. \cite{Fontanini2017}}.
\label{Fig:PF_const}
\vspace*{-1em}
\end{figure}

The P-F operator($P$) is defined in the discrete representation of the space $X$. The space, $X$, is discretized into a $D_k$ cells/states for $k=1,\ldots, N$. In this finite dimension discrete setting the time-evolution is given by the linear system in Eq.(\ref{Eq:3}) .
\begin{eqnarray}
&&\mu_{t_{i+1}} = \mu_{t_{i}} P + \hat{S}_{t_{i},t_{i+1}} \hspace{2 mm} i \in \{0,\dots,m \} \nonumber\\
&&y=\mu_{t_i}C=\mu_{t_i}[c_1\ldots c_p]
\label{Eq:3}
\end{eqnarray}
with $\mu_{t_i}\in \mathbb{R}^N$ and $c_i$ are the individual columns of matrix $C$. The source term, $\hat{S}_{t_i,t_{i+1}}$, includes volumetric and inlet sources in the domain \cite{Fontanini2017} and the matrix $C\in\mathbb{R}^{N\times p}$ with each column consisting of either 0 or 1 with one corresponding to location of sensor and zero everywhere.\\
The $\mu_{t_{i}}$ is the discrete analog to the scalar field at the given time $t_{i}$, as the cell volumetric average of $\Phi$ given by
\begin{equation}
\mu_{t_{i}}(D_k) = \frac{1}{V_{D_{k}}} \int_{D_{k}} \Phi(x,y,z,t_{i}) dV, {D_k} =1,\cdots,N 
\end{equation}
In the Markov matrix $P$, each row $i$ represents where the state $\omega_{k}$ would transition to in the next Markov time-step. To find these transition values a concentration of 1.0 is placed only at state $D_{k}$ and then computing how the initial concentration will spread to the rest of the states during the Markov time step,Eq.(\ref{Eq:MarkovConc}). 
\begin{equation}
P_{(D_{k},D_{k}+1)}(i,j) = \mu_{t_{i}(D_{k}+1)}(j) \qquad i= 1:N
\label{Eq:MarkovConc}
\end{equation}
To find more details interested readers can look at the work of Fontanini et.al \cite{Fontanini2017}. Figure \ref{Fig:PF_const} shows the process of constructing each row of the $P$ matrix. Once the $P$ matrix is constructed, we can propagate any initial contaminant distribution in time very efficiently \cite{Fontanini2017}.

\subsection{Observability gramian and optimal sensor placement} \label{sec:ContamiantTrack}
Once we have the Markov matrix, it can be used to create the contaminant history for a longer time period. This is called as the contaminant tracking matrix $\mathbf{Q}_{\tau}$ and is defined as follows \cite{Fontanini2016}
\begin{equation}
\mathbf{Q}_{\tau} = I + P + P^2 +\dots+ P^m.
\label{Eq:4}
\end{equation}
\begin{figure}
\centering
\includegraphics[scale=0.7]{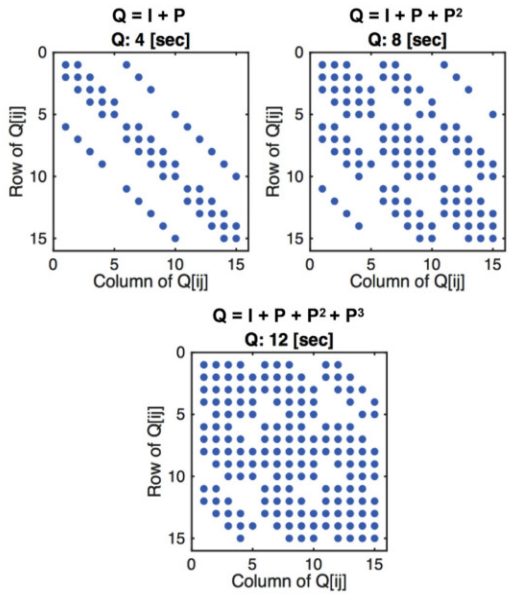}
\caption{The contaminant tracking matrix calculated from Eq.(\ref{Eq:4}) for 4,8 and 12 sec of an aircraft cabin \cite{Chen2014}}
\label{Fig:Chen Qmatrices}
\end{figure}
Each row of the contaminant tracking matrix contains the information about the transport history of a constant contaminant source at one cell (i.e one row). In case of steady state flow field the contaminant tracking matrix is given by Eq.(\ref{Eq:4}).
Since $P^m( = P \times P \times \dots \times P  \text{ m times})$ is the $m^{th}$ multistep transition matrix, the contaminant tracking matrix builds a history as to where the contaminant will be propagated in $m$ Markov time-steps. To illustrate this Fig. \ref{Fig:Chen Qmatrices} 
shows a graphical example of this process. It can be observed that as the matrix is propagated in time it get denser and the entries getting filled are the states where the contaminant will be present.

The system theoretic concept of observability gramian can be extended to nonlinear system using  transfer P-F operator \cite{vaidya2007observability}. We used for the purpose of optimal sensor placement in advection diffusion PDE in \cite{vaidya2012actuator}. This notion of observability can now be applied to Eq.(\ref{Eq:3})  to define relative degree of observability of various sets $D_k$ in state space for the given configuration of the sensor locations on set $A_k$. In particular, the relative degree of observability for cell $D_k$ over time period  $[0,\tau]$ with sensor configuration as defined by matrix $C$ is given by
\[{\cal O}_k=e_k^\top {\bf Q
}(c_1+c_2+\ldots +c_p)\]
where $e_k$ is a column vector of all zeros except for one at location $k$. Note that ${\cal O}_k$ is scalar and if ${\cal O}_k>0$ then contaminant with support on cell $D_k$ is observable with sensor configuration $C$. Furthermore, if ${\cal O}_k> {\cal O}_\ell>0$ for $k\neq \ell$ then
 cell $D_k$ is more observable than cell $D_\ell$. This relative degree of observability criteria can now be used in the design of sensor configuration matrix $C$ to achieve maximum degree of observability for all the cell $D_k\subset X$. For finding the sensor location, consider $a^{th}$ release location in the Markov states as $r^{(a)}$ given as.
 \begin{equation}
 r^{(a)} \in \mathbb{R}^{1\times N}, r_{i}^{(a)} = \begin{cases}
    1 ,& \text{i}=\text{a}\\
    0, & \text{otherwise}
\end{cases} 
\end{equation}
 For including all the release scenarios of the Markov state, the release scenario matrix $r^{(all)} = \bf I$. Therefore, the sensor location should be chosen to maximize all the observable states and the release scenarios. The problem is equivalent to the combinatorics problem of "set cover" (which in an NP hard problem), and therefore a nearly optimal greedy algorithm \cite{chvatal1979greedy}  is used. The $1^{st}$ sensor is placed at the cell corresponding to where the column support is maximized. This results in some fraction of locations which are observed by the first sensor. For the next sensor locations, these states are removed from the contaminant tracking matrix to result in $\hat{\bf Q}$. The matrix is updated for every new sensor placement to reflect unobserved states. The next sensor is placed based on the set of remaining unobserved states as release locations $r^{(B)}$. The complete algorithm-\ref{Alg:the_alg} is outlined for optimal placement of sensors.
 \begin{algorithm}
 \caption{Algorithm for Sensor Placement}
 \label{Alg:the_alg}
 \begin{algorithmic}[1]
 \renewcommand{\algorithmicrequire}{\textbf{Input:}}
 \renewcommand{\algorithmicensure}{\textbf{Output:}}
 \REQUIRE $\mathbf{Q}$, No. of sensors($p$)
 \ENSURE  $l_{p} \in X $ \\
  \STATE $l_{1} = \underset{j}{ \max } \big \{  \big[\bf I \bf Q \big]_{(:j)}  \big \}$
  \FOR {$i = 2 $ to p}
  \STATE $\bf \hat{Q}$: States removed based on $l_{i-1}$
  \STATE $r^{(B)}$ : Remaining unobservable release locations
  \STATE $l_{(i)} = \underset{j}{ \max } \big \{ \big[ r^{(B)} \bf \hat{Q} \big]_{(:j)}  \big \}$
  \ENDFOR
 \RETURN $l_{p}$ 
 \end{algorithmic} 
 \end{algorithm}
The positivity property of Markov matrix is used to show that the above greedy algorithm leads to the optimal location of sensors \cite{sinha2013optimal,sinha2016operator}. For more details, the readers are recommended to look in \cite{Fontanini2016}. While the above algorithm describes systematic procedure for optimal placement under ideal conditions, the following practical consideration has to be taken into account before applying the sensor placement algorithm. 


\subsubsection{Thresholding \& Incorporating Placement Constraints}\label{sec:thresholding}
Every sensor has an associated sensing accuracy which must be taken into account while designing a sensor network. The accuracy threshold depends on the quality of the sensor. The threshold value is a non-dimensional value and is described as the ratio of minimal value detected to the source release rate $S_{source}$,the release time (sensing/response time) $\tau$, $\epsilon_{acc} = \frac{\mu_{detect}}{S_{source} \tau} = \frac{\mu_{detect}}{\mu_{source}}$ \cite{Fontanini2016}. The values $\epsilon_{acc}$ are usually prescribed by the sensor manufacturer. After construction of ${\mathbf{Q}}_{\tau}$, we can apply this threshold as Eq.(\ref{Eq:5}).
\begin{equation}
{\mathbf{Q}^{*}_{\tau}} = {\mathbf{Q}}_{\tau} > \epsilon_{acc} 
\label{Eq:5}
\end{equation}
Eq.(\ref{Eq:5}) can be looked as an operator to convert a real valued matrix to a binary entry matrix $\mathbf{Q}^{*}_{\tau}$. Elements in the matrix that contain value one correspond to the states which are sensed by the sensor based on the accuracy threshold. \\
\indent Another important aspect while considering the sensor location is the suitability of that location. The sensor cannot be placed in the occupied zone in the building as it can hinder the occupant's movement. Furthermore, there are aesthetic and design limitations which also need to be accounted for.  To account for the placement location constraint, if a state is not able to accommodate the sensor, then the column corresponding to it is removed and replaced by zeros. Suppose $j \in \mathbb{N}^{nloc}$ the entries of $j^{th}$ in the $\mathbf{Q}^{*}_{\tau}(:,j) = 0$.\\ 
\indent In some particular applications like indoor air quality(IAQ) where $CO_{2}$ monitoring is required in the occupied space. The sensor should be placed such that it only monitors the occupied region. In this case, the states which are not in the occupied space are removed. To remove these states $i\in\mathbb{N}^{nsen}$ for the sensor placement the $i^{th}$ row is replaced by zeros, i.e $\mathbf{Q}^{*}_{\tau}(i,:) = 0$.
\section{Extension of results to uncertain operating conditions}\label{sec:UncertainOper}
Until now we discussed the sensor placement approach developed for a deterministic flow field, now we will extend the formulation under the uncertain conditions. 
\subsection{Problem Definition in Stochastic Space}
Consider the domain $\Omega \subset \mathbf{R}^d$ with the boundaries $\partial \Omega$, in which sensor locations have to be found. To define uncertainty for the problem we define a complete probability space  $(\bf D,\mathcal{F},\mathcal{P})$ where $\bf D$ is the event space, $\mathcal{F}$ is the $\sigma$-algebra and $\mathcal{P}: \mathcal{F}\rightarrow [0,1]$ is the probability measure. We then define a set of random variables (occupant location) $\widetilde{\pmb{\xi}} = \{ \pmb{\xi}: \pmb{\xi} \in \mathbf{R}^d \} \in \mathbf{R}^{\bf{K}}$ on the probability space. $\bf{K}$ being dimensions of the random set. The set consists of random variables $\pmb{\xi}$ which affects the flow field in the domain $\Omega$. In case large $\bf{K}$ it is impossible to simulate all the conditions and hence a sampling is carried from the random variable set $\widetilde{\pmb{\xi}}$ to chose $M$ samples. The set $\mathbf{S} = \{s_{1},\dots,s_{M}\}$ are chosen such that they represent the $\widetilde{\pmb{\xi}}$ distribution in a statistical sense. Each sample has an associated weight represented as $\pmb{\Theta}=\{\theta_{1},\dots,\theta_{m}\}$. Now, the objective is to find the sensor locations in $\Omega \times \bf D$, to obtain the probabilistic coverage of the domain $\Omega$ with the optimal sensor location.
\subsection{Construction of Volume Coverage in Expectation Sense}
\begin{figure}
\centering
\includegraphics[scale=0.5,width=0.95\linewidth,height=0.5\linewidth]{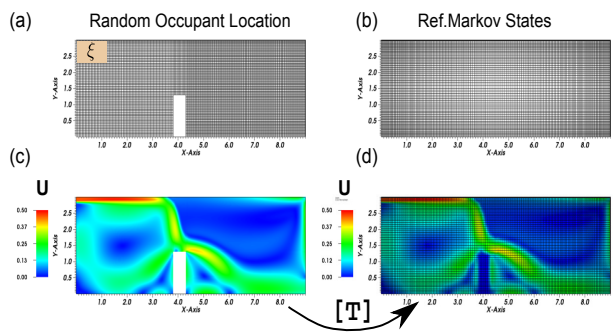}
\caption{(a-b) The occupancy Markov state mesh and the reference Markov mesh used for mapping the random occupant location flow field on to it. (c-d) The velocity field of obstruction case  which is mapped on to the reference state case using mapping matrix $[T]$}
\label{Fig:Meshesmap}
\end{figure}

Each sampling point will have a flow realization associated with it, therefore we can write the set as $ \mathbf{U} = \big \{U_{1},\dots,U_{M}\big\}$. In case the $M$ sample points are chosen uniformly from the space, each realization will have same weight i.e $\pmb{\Theta}= \frac{1}{M}$. For more informed sampling one will require assuming a probability distribution function associated with the random variable to calculate the weights. Now, with the set of flow realization associated with each random sample, a set of P-F operator can be constructed as discussed in section-\ref{sec:ConstructPF}. An important thing to notice here is that the number of states in the transition matrix $P \in \mathbf{R}^{N \times N}$ is equal to the number of cells($N$) in the discretization \cite{Fontanini2017}. Therefore, for each sample, the size of the $P$ matrix will be different as each occupancy is modeled as a separate problem.
For example, Fig.\ref{Fig:Meshesmap}(a) which represent $\xi_{1}$ realization has $(p\times q)$ number of states and similarly some other location $\xi_{2}$ can have  $(l\times b)$ due to the different location of obstructions. Therefore, the expectation operator cannot be applied directly with such realizations.\\
\indent To overcome this issue a reference Markov discretization Fig.\ref{Fig:Meshesmap}(b) with fixed number of states is constructed and the flow fields $\mathbb{U}$ are mapped onto the reference state. An example for a realization with an obstruction present is shown in the Fig.\ref{Fig:Meshesmap}(c), where we map the velocity field of the obstruction states onto the reference states. We ensure that the cells in the obstruction region have zero velocity as shown in Fig.\ref{Fig:Meshesmap}(d). The mapping results in the set $\mathbf{\widetilde{U}}=\{ \widetilde{U}_1,\dots,\widetilde{U}_M \}$ on the reference grid and then the set  $\mathbf{\widetilde{P}}=\{\widetilde{P}_1,\dots,\widetilde{P}_M \}$ is constructed, where each $\widetilde{P}_{i}$ are constructed using the method discussed in section-\ref{sec:ConstructPF}. From $\widetilde{P}$ the contaminant tracking history is generated to construct the set of of contaminant tracking matrices $\{\mathbf{Q}_{\tau} = Q_{\tau,1},\dots,Q_{\tau,M}\}$. The thresholding as discussed in section-\ref{sec:thresholding} (Eq.(\ref{Eq:5})) is applied on $\mathbf{Q}_{\tau}$ to result in $\mathbf{\widetilde{Q^*_{\tau}}}=\{\widetilde{Q^*_{\tau,1}},\dots,\widetilde{Q^*_{\tau,M}}\}$.\\
\indent  Next, we compute the volumetric coverages for each flow realization. To compute this we multiply the constructed set  $\mathbf{\widetilde{Q^*_{\tau}}}$ with  $\textbf{V}$, where $\textbf{V}$ is a diagonal matrix with diagonal entries as the normalized cell volumes,  $V_{i,i}= V_{\omega_i}/ V_{tot}$. In case of a uniform discretization $\textbf{V}= \mathbf{I}$. The operation results in another set $\widetilde{\mathbf{Q}^{**}}$ Eq.(\ref{Eq:7}).
\begin{equation}
\widetilde{\mathbf{Q}^{**}} = \{ \widetilde{Q^*_{\tau,i}} \cdot \textbf{V}\}, \hspace{2mm} i \in \{1,\dots,M \}
\label{Eq:7}
\end{equation}

\subsection{Sensor Placement and Computation of Probable coverage}
\begin{figure}[h]
\centering
\includegraphics[scale=0.5,width=0.95\linewidth,height=0.4\linewidth]{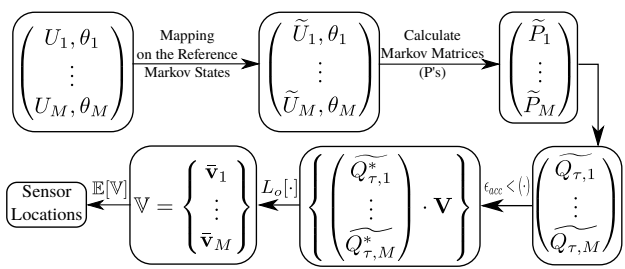}
\caption{The flowchart of the steps to account the uncertainty to find the sensor location.}
\label{Fig:flowUncertain}
\end{figure}

To compute the sensor locations $\widetilde{\mathbf{Q}^{**}}$ is used. For each matrix in this set we find the column with the maximum support by using the \textit{greedy algorithm}. This result into a collection of columns as a set $\mathbb{V} = \{\mathbf{\bar{v}}_1, \dots,\mathbf{\bar{v}}_M\}$, where each element vector $(\mathbf{\bar{v}}) \in \mathbf{R}^{n}$, $n$ being the number of states/cells in the domain. This can be compactly represented as  Eq.(\ref{Eq:8}), with $L_{o}$ operator as a greedy algorithm which returns the column with maximum support.
\begin{equation}
\begin{split}
\mathbb{V} = L_{o} \big\{ \widetilde{\mathbf{Q}^{**}} \big\}  \quad i = 1,\dots,M \\
 \end{split}
\label{Eq:8}
\end{equation}
The next step is to calculate \textbf{\textit{expectation coverage}} by associating $\bar{\mathbf{v}}_{M}$ with the weights $\theta_{M}$. The expected $\mathbb{E}[\bar{\mathbb{V}}]$ is obtained using Eq.(\ref{Eq:9a}). To find the location of the first sensor location, the index of the maximum entry in the expected vector $\mathbb{E}[{\mathbb{V}}]$ is used to find the corresponding coordinates in the domain Eq.(\ref{Eq:9b}). To find the next sensor, the information of the first sensor is removed from consideration, by removing the location for each element of the set $\widetilde{\mathbf{Q}^{**}}$ and repeating Eq.(\ref{Eq:8}-\ref{Eq:9b}). The process is repeated till all the sensors are placed in the domain.
\begin{subequations}
\begin{equation}
\mathbb{E}[{\mathbb{V}}] = \sum_{i=1}^{M} \Theta_{i}\bar{\mathbf{v}_{i}} 
\label{Eq:9a}
\end{equation}
\begin{equation}
\bar{k}(1) = max(\mathbb{E}[\bar{\mathbb{V}}])
\label{Eq:9b}
\end{equation}
\end{subequations}
 Figure-\ref{Fig:flowUncertain} discuss the overview of the complete algorithm as a flow chart. The flow fields constructed on different mesh/number of states are first mapped onto a reference number of Markov state. Each realization is associated with a probability. For each mapped flow realization the associated Markov matrices are computed. The matrices are then use to construct the contaminant matrices where the final time $\tau$ is decided by the sampling rate of the sensor. The next step is to use the constraints associated with the sensor accuracy or location. Once the thresholded set is constructed we apply the greedy algorithm to find the set of coverage vectors. The expectation operator is then applied on this set based on there associated weights. The expectation coverage vector is then used to find the index of the maximum entry in the vector to place the sensor. In the next section the results obtained for deterministic sensor placement and incorporating uncertainty are discussed in detail.\\

\section{Results}\label{sec:Results}
\begin{figure}[h]
\centering
\includegraphics[scale=0.5,width=0.95\linewidth]{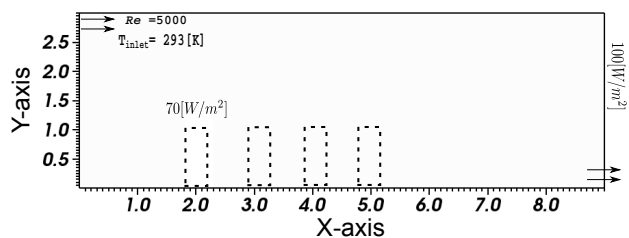}
\caption{The IEA-Annex 2D geometry with the boundary conditions and the occupancy locations used in the current study to demonstrate the framework.}
\label{Fig:GeoProb}
\end{figure}

\begin{figure}
\centering
\includegraphics[scale=0.5]{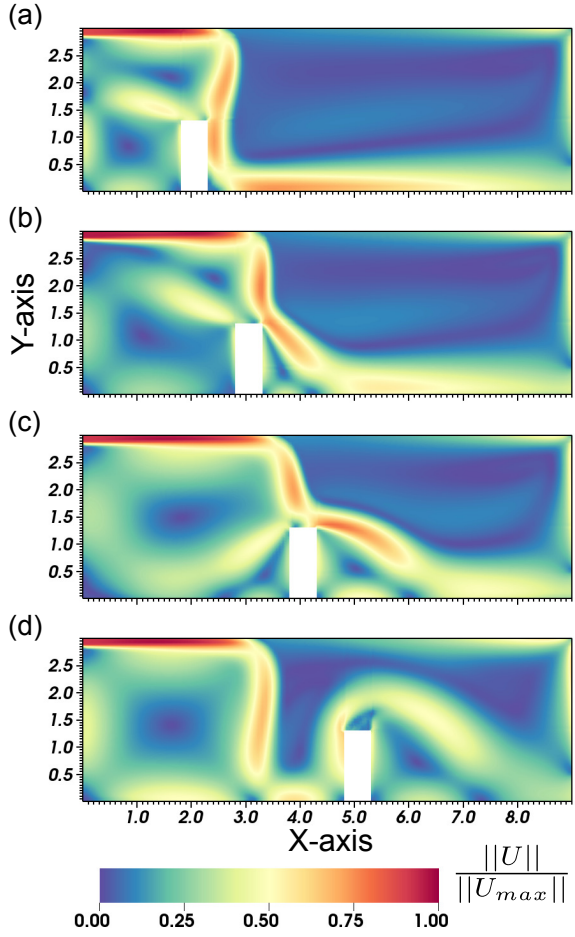}
\caption{The flow field for the four occupancy computed using CFD for different locations considered in the current work for sensor placement.}
\label{Fig:FlowField}
\end{figure}

To present the results, we take the IEA-annex 2D benchmark problem by \cite{Nielsena} with its associated dimensions. To account for the occupancy uncertainty, four positions are considered in the occupied zone of the building. The geometry consists of one inlet on the left wall and one outlet on the right wall, shown in Fig-\ref{Fig:GeoProb}. We used Reynolds number of 5000 and temperature of 293K as the boundary condition. The person is modeled as a heat source generating 70 W/$m^2$. The right heated wall in maintained at 100 W/$m^2$, while the other boundaries are insulating, all the walls were enforced with a  no-slip boundary condition. An open source finite volume code OpenFOAM \cite{Jasak2007} is used to compute the flow fields. We ran each case to the tolerance of $10^{-7}$, and a rigorous validation was carried out by comparing the non-isothermal results with the experimental data as well. The computed flow fields are shown in Fig-\ref{Fig:FlowField}, it can be seen that occupancy significantly affected the flow field as the fields are quite different. The Markov maps were then constructed for each case. They were then validated for the contaminant transport before constructing the contaminant tracking matrix, which is discussed in the next section.

\subsection{Validation of Contaminant Transport by Transfer Operator.}
\begin{figure}
\centering
\includegraphics[scale=0.5,width=0.98\linewidth]{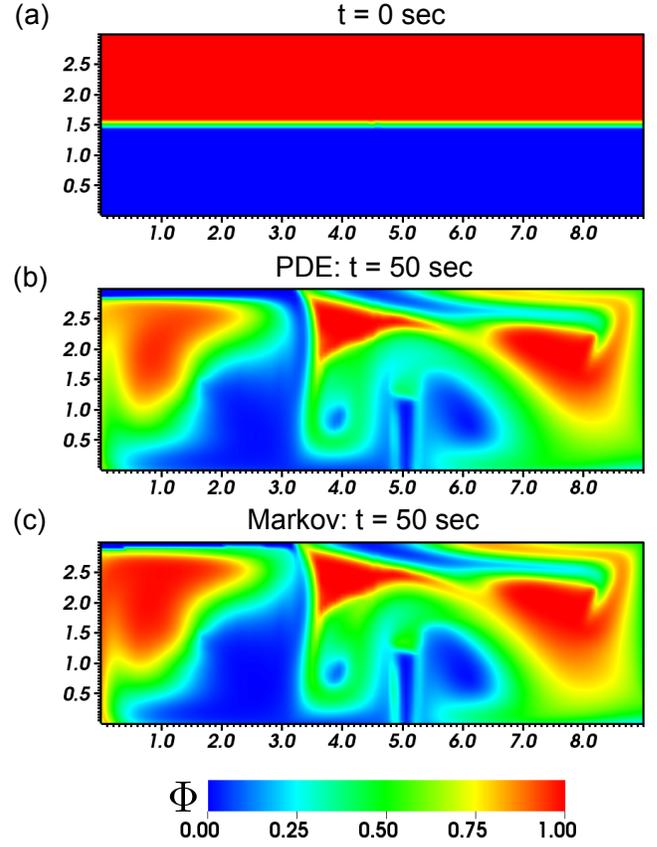}
\caption{The comparison of the concentration evolution using PDE transport and Markov matrix for a same initial concentration.(a) Initial condition (b) PDE evolution at $t= 50 \text{sec}$ (c) Markov evolution at $t = 50 \text{sec}$  }
\label{Fig:ConcenCompare}
\end{figure}

We here show the validation of the contaminant transport based on  Markov matrix for one flow-field realization (Fig-\ref{Fig:FlowField}(d)). The three remaining realizations are also validated before constructing the contaminant transport matrix. The contaminant (passive scalar) is initialized in half of the building domain at $t=0$ as shown in Fig.\ref{Fig:ConcenCompare}(a). The Markov matrix constructed with the flow field information is then used to advect this concentration to $t = 50 \text{sec}$ by matrix-vector multiplication. To compare the Markov advection results the same initial condition is also advected using the scalar transport equation-\ref{Eq:1}. Figure-\ref{Fig:ConcenCompare}(b-c) compare the advection maps of the scalar. It can be seen that the two contours are nearly indistinguishable. This validation shows the accuracy of the Markov approach and the effectiveness of simple matrix-vector product based approach for scalar transport prediction. Further, these matrices can now be used for generating the contaminant history which will be then used for the sensor placement. We note that the PF based results took x10 less time than the PDE based result.

\subsection{Sensor Placement Under Deterministic Setting }
\begin{figure}[h]
\centering
\includegraphics[scale=0.5]{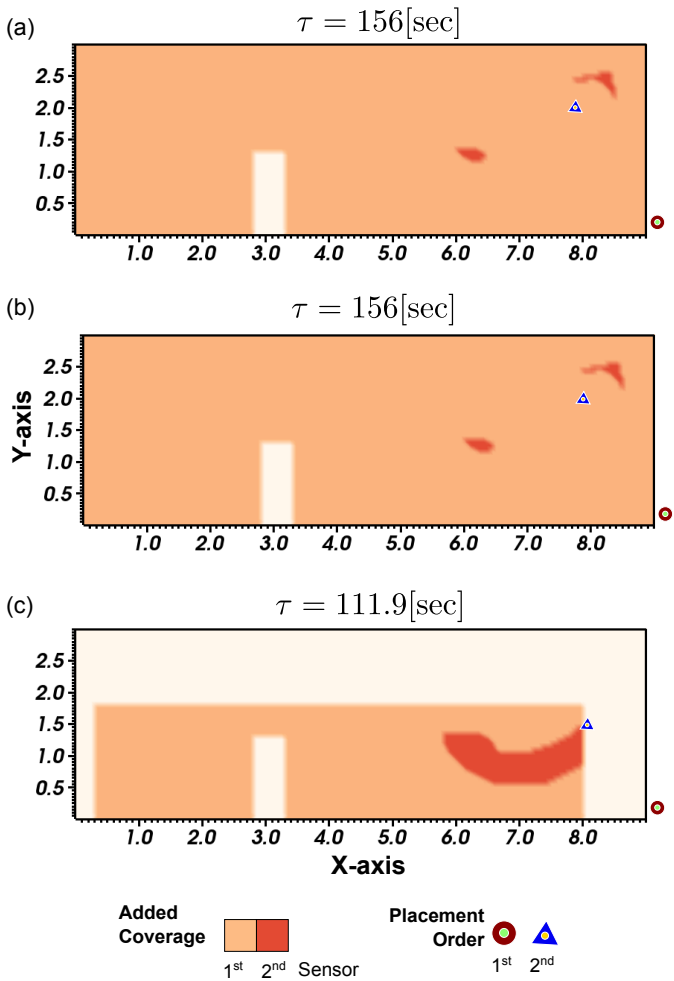}
\caption{Deterministic sensor placement for the flow field (Fig-\ref{Fig:FlowField}(b)) under (a) No-constraint (b) Location constraint (c) Sensing and Location constraints}
\label{Fig:deterCoverall}
\end{figure}

We first present the sensor placement algorithm results in the deterministic case where we use the Markov matrix for the flow field Fig-\ref{Fig:FlowField}(b). We present results for two sensor locations under three placement scenarios namely; (1) no-constraint, where the sensors can be placed anywhere in the domain, (2) location constraint, where the sensor is required to be placed outside the occupied region in the domain, (3) sensing and location constraint, where the sensor is required to be placed outside the domain but should only detect the occupied region. Figure-\ref{Fig:deterCoverall} (a-c) shows these cases as the added coverage map of the two sensors and the order in which they are placed in the domain. It can be seen that the first sensor location  turns out to be outside domain since based on the chosen final $\tau$, the contaminant will end-up at outlet irrespective of release location because of the flow field. Therefore, the algorithm returns a location which is at the outlet. Furthermore, the coverage for the outlet sensor location is the maximum and the second sensor is covering some portion of the domain. Another, observation which is important to notice is that the sensing constraint and location constraints are respected by the algorithm which are also shown in the results.

\subsection{Sensor placement accounting for Uncertainty}
\begin{figure}
\centering
\includegraphics[scale=0.4,width=0.8\linewidth]{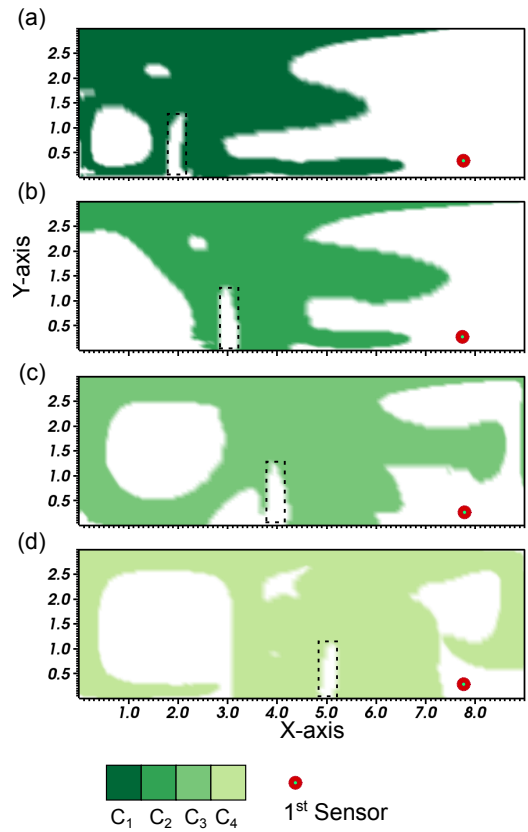}
\caption{The coverage map of a single sensor corresponding to considered four realization.}
\label{Fig:IndividualCover}
\end{figure}

\begin{figure}
\centering
\includegraphics[scale=0.5,width=0.9\linewidth]{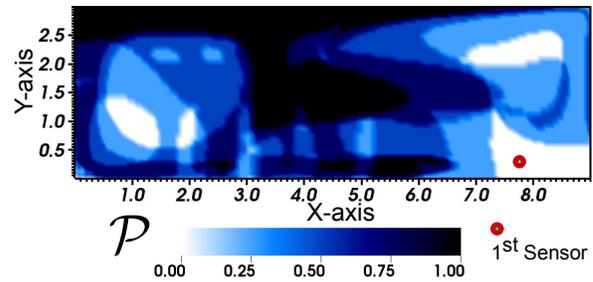}
\caption{The probability coverage map by combining individual coverages with there associated weights}
\label{Fig:ProbCover}
\end{figure}

\begin{figure}
\centering
\includegraphics[scale=0.5,width=0.9\linewidth]{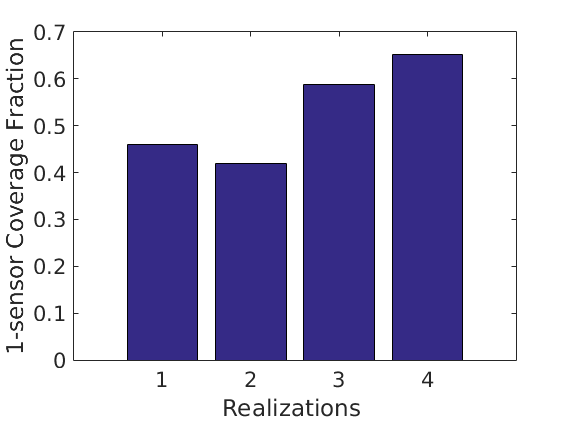}
\caption{The fraction of domain covered by the sensor placement for the individual realization, four in this case.}
\label{Fig:FracCover}
\end{figure}

The results for the sensor placement under the uncertain case of four flow realization with no-constraint on the placement location is considered. we set $\tau = 80 sec$ to construct the contaminant matrix set. To account for the sensor accuracy $\epsilon_{acc} = 0.01\%$ is used, and the weights associated with each realization is taken as $\pmb{\Theta} = 0.25$. The sensor location is then found using the discussed algorithm. The volume coverage related to each realization corresponding to the proposed algorithm is shown in Fig.-\ref{Fig:IndividualCover}. The sensor predicted by the algorithm is close to the outlet due to the long time horizon for the given flow field. Another important observation is that the choice of Markov matrix time step $\Delta t$ (set to$\Delta t  = 10 sec$) results in the volume around the sensor remaining unobserved. 
\indent Further, a more informative result can be computed with these individual coverage maps, by weighting these map by their associative weights to result in a probable coverage map. Figure-\ref{Fig:ProbCover}, shows the likely coverage contours ($\mathcal{P}$). The white regions in the contours are those regions which are less likely to be observed by the placed sensor location, and the darker overlapped area represent the probabilities by which they can be observed by the sensor, based on the setup of the problem. The coverage associated with each realization can also be calculated for the computed sensor location. Figure-\ref{Fig:FracCover}, shows the fraction of domain covered for the four realizations taken in the study. Based on this the \textit{expected coverage} is computed, which is equal to 52.96\% for the sensor location in the domain. 

\section{CONCLUSIONS}
The paper extends the dynamical system approach to designing an optimal sensor placements under uncertain conditions. The presented approach demonstrates the use of observability ideas from the control theory to develop a framework accounting uncertainty associated with the occupancy of the building. The method results in a probable coverage map which is useful in understanding the coverage behavior in multiple scenarios. The method uses the airflow information to construct the sensor location map and developed in such a way to account various sensor placement constraints. The work lays the foundation for extending the approach to complex 3D building geometries. The stochastic framework developed in this work can be easily extended to other uncertainties which can affect the airflow inside the buildings such as weather conditions, HVAC operation, etc. The dynamical system framework used in the method makes it easier to extend the approach to source estimation problem and source identification problem which will be presented as a future work.  

\addtolength{\textheight}{-1cm}   




\section*{ACKNOWLEDGMENT}
UV will like to acknowledge the financial support from National Science Foundation CAREER grant ECCS  1150405. BG acknowledges partial support from NSF CAREER 1149365.

\Urlmuskip=0mu plus 1mu\relax
\bibliographystyle{IEEEtran}
\bibliography{references}

\end{document}